\documentclass[sigconf,screen]{acmart}

\makeatletter
\def\@ACM@checkaffil{% Only warnings
    \if@ACM@instpresent\else
    \ClassWarningNoLine{\@classname}{No institution present for an affiliation}%
    \fi
    \if@ACM@citypresent\else
    \ClassWarningNoLine{\@classname}{No city present for an affiliation}%
    \fi
    \if@ACM@countrypresent\else
        \ClassWarningNoLine{\@classname}{No country present for an affiliation}%
    \fi
}
\makeatother

\AtBeginDocument{%
  }

\usepackage{algorithm}
\usepackage[noend]{algpseudocode}

\usepackage{enumitem}

\usepackage{listings}
\definecolor{darkgreen}{rgb}{0, 0.44, 0.23}
\definecolor{lightgreen}{rgb}{0.25, 0.63, 0.4375}
\definecolor{darkblue}{rgb}{0.02, 0.16, 0.49}
\usepackage{microtype}

\usepackage{multirow}

\newcommand{\tool}{{\textsc{SeeWasm}}}

\setcopyright{acmlicensed}
\acmDOI{10.1145/3650212.3685300}
\acmYear{2024}
\copyrightyear{2024}
\acmISBN{979-8-4007-0612-7/24/09}
\acmConference[ISSTA '24]{Proceedings of the 33rd ACM SIGSOFT International Symposium on Software Testing and Analysis}{September 16--20, 2024}{Vienna, Austria}
\acmBooktitle{Proceedings of the 33rd ACM SIGSOFT International Symposium on Software Testing and Analysis (ISSTA '24), September 16--20, 2024, Vienna, Austria}
\acmSubmissionID{issta24demo-p4-p}
\received{2024-07-05}
\received[accepted]{2024-07-26}

\begin{document}

% FOR ALIGN* ENVIRONMENT
\setlength{\abovedisplayskip}{2pt}
\setlength{\belowdisplayskip}{2pt}
\setlength{\abovedisplayshortskip}{2pt}
\setlength{\belowdisplayshortskip}{2pt}

\title{SeeWasm: An Efficient and Fully-Functional Symbolic Execution Engine for WebAssembly Binaries}

\author{Ningyu He}
\orcid{0000-0002-9980-7298}
\affiliation{%
  \institution{Key Lab of HCST (PKU), MOE; SCS, Peking University}
  \city{Beijing}
  \country{China}
}
\email{ningyu.he@pku.edu.cn}

\author{Zhehao Zhao}
\orcid{0000-0001-6975-8352}
\affiliation{%
  \institution{Key Lab of HCST (PKU), MOE; SCS, Peking University}
  \city{Beijing}
  \country{China}
}
\email{zhaozhehao@pku.edu.cn}

\author{Hanqin Guan}
\orcid{0009-0008-3723-4915}
\affiliation{%
  \institution{Key Lab of HCST (PKU), MOE; SCS, Peking University}
  \city{Beijing}
  \country{China}
}
\email{hqguan@stu.pku.edu.cn}

\author{Jikai Wang}
\orcid{0009-0001-0887-5560}
\affiliation{%
  \institution{Huazhong University of Science and Technology}
  \city{Wuhan}
  \country{China}
}
\email{wangjikai@hust.edu.cn}

\author{Shuo Peng}
\orcid{0009-0003-9141-3221}
\affiliation{%
  \institution{Huazhong University of Science and Technology}
  \city{Wuhan}
  \country{China}
}
\email{1874834431@qq.com}

\author{Ding Li}
\orcid{0000-0001-7558-9137}
\affiliation{%
  \institution{Key Lab of HCST (PKU), MOE; SCS, Peking University}
  \city{Beijing}
  \country{China}
}
\email{ding_li@pku.edu.cn}

\author{Haoyu Wang}
\orcid{0000-0003-1100-8633}
\affiliation{%
  \institution{Huazhong University of Science and Technology}
  \city{Wuhan}
  \country{China}
}
\email{haoyuwang@hust.edu.cn}

\author{Xiangqun Chen}
\orcid{0000-0002-7366-5906}
\affiliation{%
  \institution{Key Lab of HCST (PKU), MOE; SCS, Peking University}
  \city{Beijing}
  \country{China}
}
\email{cherry@sei.pku.edu.cn}

\author{Yao Guo}
\orcid{0000-0001-5064-5286}
\affiliation{%
  \institution{Key Lab of HCST (PKU), MOE; SCS, Peking University}
  \city{Beijing}
  \country{China}
}
\email{yaoguo@pku.edu.cn}

% \renewcommand{\shortauthors}{Ningyu et al.}

\begin{abstract}
WebAssembly (Wasm), as a compact, fast, and isolation-guaranteed binary format, can be compiled from more than 40 high-level programming languages. However, vulnerabilities in Wasm binaries could lead to sensitive data leakage and even threaten their hosting environments.
To identify them, symbolic execution is widely adopted due to its soundness and the ability to automatically generate exploitations.
However, existing symbolic executors for Wasm binaries are typically platform-specific, which means that they cannot support all Wasm features. They may also require significant manual interventions to complete the analysis and suffer from efficiency issues as well.
In this paper, we propose an efficient and fully-functional symbolic execution engine, named {\tool}.
Compared with existing tools, we demonstrate that {\tool} supports full-featured Wasm binaries without further manual intervention, while accelerating the analysis by 2 to 6 times. {\tool} has been adopted by existing works to identify more than 30 0-day vulnerabilities or security issues in well-known C, Go, and SGX applications after compiling them to Wasm binaries.
\end{abstract}




\maketitle

\section{Introduction}
\label{sec:intro}
WebAssembly (Wasm) was introduced and proposed in 2017~\cite{haas2017bringing}.
As a static-type stack-based binary format, Wasm can serve as an \textit{efficient} and \textit{isolation-guaranteed} compilation target. 
\textit{Versatility} is one of the significant advantages of Wasm. On one hand, more than 40 compilation toolchains take Wasm as their compilation target~\cite{40lang}, \textit{e.g.}, \texttt{clang} for C/C++ and \texttt{rustc} for Rust. On the other hand, web browsers and standalone runtimes allow running Wasm binaries on different operating systems and even hardware architectures.
Developers have been adopting Wasm in production environments. For example, PhotoShop and AutoCAD can be executed in web browsers in the Wasm format~\cite{wasm-projects}, while SQLite and Nginx can be run in Wasm standalone runtimes on edge-side devices~\cite{sqlite}.

Though Wasm runtimes guarantee the isolation between Wasm binaries and their hosting environments, Wasm binaries themselves may suffer from security threats.
For example, Lehmann \textit{et al.} illustrate that vulnerabilities in source code could still be exploitable in Wasm binaries, which can even affect their hosting environment~\cite{lehmann2020everything}.
Attackers can obtain attack primitives and exploit hosting environments by exploiting vulnerabilities in Wasm binaries.

Symbolic execution, as one of the representative static analysis methods, has been applied to detect vulnerabilities in Wasm binaries.
For example, WANA~\cite{jiang2021wana} and EOSafe~\cite{he2021eosafe} both target EOSIO smart contracts written in Wasm and have successfully identified various types of critical security issues.
Manticore~\cite{mossberg2019manticore}, as an open-source commercial project (more than 3.7K stars on GitHub~\cite{manticore}), is an actively maintained symbolic executor for Wasm binaries.

However, current tools are restricted by either \textit{limited functionality} or \textit{low efficiency}.
Specifically, some tools ignore the constraints that instructions should follow or do not support imported functions, \textit{i.e.,} \textit{partially-functional}. This is for robustness reasons, \textit{i.e.,} to minimize the possibility of tools crashing due to unexpected behaviors when analyzing binaries.
For example, Manticore cannot handle import functions, whose behaviors must be provided by users via hooking functions in separate test drivers or writing binary instructions directly.
As for the \textit{low efficiency}, current tools do not integrate general or Wasm-specific optimization strategies, but simply exchange for efficiency by adapting to specific platforms.
For example, to accelerate the analysis, EOSafe only models the external dependency in EOSIO.
Moreover, against the Wasm-specific linear memory, none of the current tools proposes methods to avoid the huge overhead introduced by state forking while considering the soundness when encountering both concrete and symbolic pointers.

In this work, we propose {\tool}, \textit{an efficient and fully-functional symbolic execution engine for Wasm binaries}.
Specifically, to address the limited functionality, we emulate all instructions in Wasm 1.0 specification~\cite{wasm-specification}. The emulator considers both instruction semantics and the corresponding behavioral constraints to guarantee soundness.
Additionally, to deal with functions in WebAssembly system interface (WASI)~\cite{wasi}, and common libraries in C and Go, we propose the corresponding function modeling with considering their side effects on control flow and data structures. 
As for the low efficiency, we propose a Wasm-specific memory modeling algorithm.
It can enumerate all possible situations specified by both concrete and symbolic pointers in a state-merging manner, eliminating the huge overhead caused by state forking. 
Moreover, we adopt a set of general caching strategies for constraint solving, \textit{e.g.}, unsat core and incremental solving, reducing the overhead of requesting solutions to the back-end SMT solver each time.

\begin{figure}[t]
\centering
\includegraphics[width=\columnwidth]{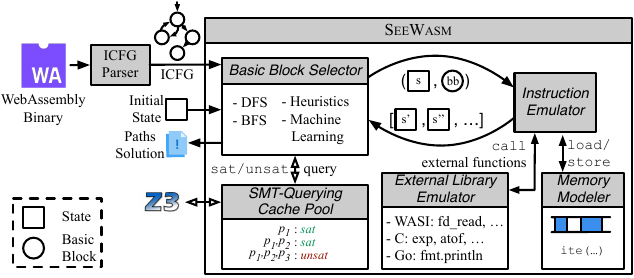}
\vspace{-0.15in}
\caption{The workflow and architecture of \tool.}
\label{fig:framework}
\vspace{-0.2in}
\end{figure}

Unlike existing tools, {\tool} completes the whole evaluation without any manual intervention. Moreover, compared to Manticore, {\tool} can improve efficiency by around 2 to 6 times. 
Such results illustrate that {\tool} can support full-featured Wasm binaries while accelerating the analyzing process.
Earlier versions of {\tool} have been adopted by \textsc{Eunomia}~\cite{he2023eunomia} and \textsc{SymGX}~\cite{wang2023symgx}, both of which take {\tool} as the core engine and detect more than 30 0-day vulnerabilities or security issues in C, Go and SGX applications after compiling to Wasm.

\section{Background}
\label{sec:background:wasm}
WebAssembly (Wasm) is a low-level assembly-like binary format, consisting of a rich set of instructions~\cite{wasm-specification}.
There are four primitive data types in Wasm, 32/64-bits integers or floats. Only primitive data types can be stored in the stack and local/global area. Other complex data types in source code, like \texttt{struct} in C, will be encoded as raw bytes and stored in the Wasm-specific linear memory area.

Besides traditional control flow instructions, \textit{e.g.}, \texttt{if-else}, Wasm supports some specific ones.
For instance, \texttt{br\_table} can simultaneously designate multiple branches and jump to the one corresponding to the dynamically given operand.
Moreover, \texttt{call\_indirect} dynamically determines the callee according to the operand as well.
These two operators can introduce more paths, possibly resulting in the \textit{path explosion} issue during the symbolic execution.

\section{Design \& Implementation}
\label{sec:design}
Figure~\ref{fig:framework} shows the workflow and architecture of {\tool}.
{\tool} takes a Wasm binary as input, and generates solutions corresponding to feasible paths and assertion failures.
We will provide a detailed description of how {\tool} addresses the limited functionality and low efficiency issues in \S\ref{sec:design:functional} and \S\ref{sec:design:efficiency}, respectively.

\subsection{Supporting Full Wasm Features}
\label{sec:design:functional}

The partial functionality of existing tools, \textit{e.g.,} Manticore, mainly because:
(1) lack of a fully-functional instruction emulator; and (2) cannot deal with import functions or their side effects.

Against the first issue, we propose \textit{behavioral constraints considered instruction emulation} in an \textit{instruction emulator} that supports all instructions defined in the Wasm 1.0 specification while considering all corresponding constraints.
Specifically, Wasm specification restricts behavioral constraints for each instruction, related to its type and semantics~\cite{wasm-specification}. Thus, for each instruction, the emulator firstly translates all behavioral constraints into assertions to ensure that no additional paths are introduced or critical paths are missed. Then, depending on whether the operand is symbolic or concrete, the emulator updates the corresponding field of the state and performs the necessary state forking.
Take the emulation of the distinctive \texttt{br\_table} instruction as an example.
The emulator firstly asserts the types of its argument and its immediate numbers. Then, if the argument is a concrete value, the control flow will be directly directed to the branch by updating the \texttt{pc} field. Otherwise, all possible branches will be enumerated by state forking.

Against the second issue, we implement \textit{side-effect-aware external dependency emulation} in the \textit{external library emulator} to handle the call to import functions.
According to their impact on the accuracy of results, two emulating strategies exist, \textit{i.e.}, \textit{side-effect-considering} and \textit{stack-balance-considering}.
The former one emulates all function behaviors precisely, including operations, \textit{i.e.,} side effects on control flow or data structures, while the latter one only guarantees the stack balance by symbolizing the return value.

To enable executing Wasm binaries on various operating systems, Wasm specifies a set of \textit{WebAssembly system interface} (WASI)~\cite{wasi} functions, which can be imported to achieve functionalities like I/O and network connection.
I/O-related WASI functions are related to symbol initialization and results export, thus we implement a \textit{symbolic file system} to handle them in side-effect-considering way.
It supports all operations on directories and files.
For properties of files, \textit{e.g.}, content and cursor, the file system abstracts each file into an object, where updating properties is equivalent to updating fields of objects.
For example, \texttt{fd\_read} accepts input from the designated file descriptor. The emulator checks if the given file descriptor is opened, retrieves bytes from the buffer, and writes them at the designated memory offset. If any of attributes is symbolic, necessary state forking will be conducted to guarantee the soundness.

Besides WASI-related functions, we also emulate widely adopted library functions in C and Go in side-effect-considering way, \textit{e.g.}, mathematical calculations.
For other uncommon import functions, the emulator symbolizes the return value and pushes it onto the stack to maximize the analysis efficiency.
Note that, we expose an interface, where users can perform emulation in the side-effect-considering way by writing Python3 codes to customize {\tool}.

\subsection{Improving Execution Efficiency}
\label{sec:design:efficiency}

For a platform-agnostic symbolic execution engine for Wasm binaries, two main reasons affect its efficiency: (1) lack of a memory modeling method specifically designed for the Wasm linear memory; and (2) no optimizations on the constraint solving process.

\vspace{-0.14in}
\begin{algorithm}[H]
\caption{State merging based linear memory modeling}\label{alg:cap}
\begin{algorithmic}[1]
\renewcommand{\algorithmicrequire}{\textbf{Input:}}
\renewcommand{\algorithmicensure}{\textbf{Output:}}
\Require $M$; $dest$; and $len$
\Ensure the \texttt{ite} statement covers all possible situations
\Procedure{LoadFrom}{$M, dest, len$}
\If{$M$ is empty}
	\State \Return \texttt{inv}
\EndIf
\State $(l, h), val \gets M.popitem()$
\State \Return \parbox[t]{200pt}{\texttt{ite}$(l <= dest <= h$,\\
\textsc{BuildIte}$(l, h, val, dest, len, 0)$,\\
\textsc{LoadFrom}$(M, dest, len))$\strut}
\EndProcedure
\Procedure{BuildIte}{$l, h, val, dest, len, offset$}
\If{$offset + len = h - l$}
	\State \Return $val[offset: offset + len]$
\Else
	\State \Return \parbox[t]{200pt}{\texttt{ite}$(l + offset = dest$,\\
$val[offset: offset + len]$,\\
\textsc{BuildIte}$(l, h, val, dest, len, offset + 1))$\strut}
\EndIf
\EndProcedure
\end{algorithmic}
\label{algo:memory}
\end{algorithm}
\vspace{-0.18in}

Specifically, for the linear memory adopted by Wasm, He \textit{et al.}~\cite{he2021eosafe} have proposed a method to model the \texttt{load} or \texttt{store} instructions when encountering concrete pointers.
However, against symbolic pointers, current methods either concretize them for efficiency or fork as many feasible states as possible for soundness.
To deal with such a dilemma, we propose a \textit{state merging based linear memory modeling algorithm} as shown in Algorithm~\ref{algo:memory}.
As we can see, through recursively constructing \texttt{if-then-else} (\texttt{ite}) statements, a single statement can enumerate all possible situations.
Specifically, the memory, denoting as $M$, is modeled by key-value mappings, where the key is a tuple to specify the address range, and the value is the corresponding data. Formally, $M[k]$ returns the corresponding data $v$, and the $v[l:h]$ means extracting from $l$ to $h$ from $v$.
The \texttt{LoadFrom($M, dest, len$)} indicates loading $len$ bytes from $dest$ in $M$. 
The algorithm returns an \texttt{ite} statement eventually, where \texttt{inv} refers to an invalid address.
Not only soundness is guaranteed through enumerating all possible situations, but also no overhead is introduced due to state copying. The \texttt{ite} query can be accelerated by the strategies as we will illustrate in the following cache pool.

KLEE~\cite{cadar2008klee} has proven that constraint solving is one of the greatest bottlenecks for symbolic execution.
Thus, we introduce a \textit{SMT-querying cache pool} to accelerate the SMT solving.
Three strategies are adopted: \textit{query cache}, \textit{unsat core} and \textit{incremental solving}.
Take a query on a set of predicates $P = [p_1, p_2, p_3]$ as an example.
\begin{enumerate}[leftmargin=15pt]
	\item The pool will check if $P$ has been queried and cached before. If it is, the result will be returned immediately;
	\item Otherwise, the pool will examine if there is an unsat core, \textit{i.e.}, whether the result of any of the subsets of $P$ is unsatisfiable. If it is, it can be guaranteed that the $P$ is unsat;
	\item Finally, the pool will try to find the \textit{maximized subset} of $P$, \textit{e.g.}, $P'$, and retrieve its solving result. Based on the solution of $P'$, the solver can find the solution of $P$ through incremental solving by adding predicates $P - P'$.
\end{enumerate}
If all three rules fail, the query will fall back to the back-end solver, and the results will be inserted into the pool.

\section{Usage Example}
\label{sec:usage}
We will take the following code snippet, a password-guessing logic bomb, as an example to show how to use {\tool}. 
\begin{lstlisting}[language=C]
int check_password(char *buf) {
  if (buf[0] == 'h' && buf[1] == 'e' && buf[2] == 'l'
      && buf[3] == 'l' && buf[4] == 'o') {return 1;}
  return 0;
}
int main(int argc, char **argv) {
  if (check_password(argv[1])) {
    printf("Password found!\n");
    return 0;
  } else { return 1; }
\end{lstlisting}
To crack the password, we start the analysis by:
\begin{align*}
\small{\texttt{python3 launcher.py -f ./pwd.wasm -s --sym\_args 10 -v info}}	
\end{align*}
, where \texttt{-s} starts the symbolic execution analysis, and \texttt{--sym\_args 10} symbolizes a 10-byte string as command line input.
Eventually, {\tool} generates possible states when encountering the line 2 condition. The one that cracks the password looks like:
% \vspace{-0.03in}
\begin{lstlisting}
"Return": "0",
"Solution": {"sym_arg_1": "hello"},
"Output": [{
        "name": "stdout",
        "output": "Password found!\n"
    },{
        "name": "stderr",
        "output": ""
    ...
\end{lstlisting}
% \vspace{-0.03in}
We can observe that the argument is solved as the password, \textit{i.e.}, \texttt{hello}. Moreover, \texttt{0} is returned and the standard output generates the \texttt{Password found!\textbackslash{}n}, \textit{i.e.}, line 8 in the source code. No error messages are captured through the standard error stream.

Other options are supported by {\tool}, like \texttt{--sym\_stdin} and \texttt{--sym\_files} can initialize stdin stream and a symbolic file system, respectively. Please refer to the documentation for more details.

\section{Evaluation}
\label{sec:evaluation}

We evaluate the usability of {\tool} both qualitatively and quantitatively, as well as its vulnerability detection ability.

\noindent
\textbf{Benchmark.}
We choose two representative benchmarks, \textit{i.e.}, b-tree test suite~\cite{marques2022concolic} ($B_1$) and Gillian-Collections-C~\cite{gillian} ($B_2$).
Specifically, cases in $B_1$ create b-tree with different degrees, manipulate it and assert the correctness of such behaviors.
$B_2$ implements a test suite for the Collections-C project~\cite{collections-c}, a well-known C library (2.8K+ stars on GitHub) consisting of 10 data structures and the corresponding helper functions.
Both of them were adopted as benchmarks to evaluate the effectiveness of related works~\cite{marques2022concolic,he2023eunomia,fragoso2020gillian}.

\noindent
\textbf{Experimental Setup.}
The selector takes the BFS strategy by default, and the back-end SMT-solver is z3 4.10.2.
Considering the maturity and influence of current symbolic executors against Wasm, we take Manticore as the baseline, which is developed and maintained by a well-known blockchain security company and open-sourced with 3.7K+ stars on GitHub~\cite{manticore}.
Moreover, we remove unnecessary instrumented instructions in $B_1$ as it was originally designed for evaluating the Wasm concolic executor. We compile cases in $B_2$ from C to Wasm with \texttt{clang} released in \texttt{wasi-sdk-14}~\cite{wasi-sdk}.
We pulled the latest stable version of Manticore on PyPI.
All evaluations are performed on a Mac with M1 Max and 32GB memory.

\subsection{Qualitative Analysis}
We first qualitatively compare the usability between {\tool} and Manticore from two aspects, \textit{i.e.}, \textit{design} and \textit{usage}.
Specifically, on one hand, \textit{the design of Manticore suffers from the scalability issue}. For example, Manticore does not provide a selector-like interface to fine-grained adjust the path searching strategy. Moreover, both the memory modeling and the constraints querying are implemented in a naive way.
On the other hand, \textit{using Manticore in analyzing tasks requires significant manual efforts}. Because Manticore does not support any import functions, users have to manually provide their implementations each time.
For example, if a Wasm binary accepts input through \texttt{scanf}, which takes a complex patterned string and interacts with the linear memory, there are two possible solutions. Either writing the implementation for the underlying WASI function \texttt{\_\_wasi\_fd\_read} in the Wasm binary, or changing \texttt{scanf} in the source code to a simpler one, like \texttt{getchar} that does not interact with memory, and recompiling it.
The former is almost impossible, because WASI function implementation is extremely complex, not to mention modifying binary directly. The latter changes the source code semantics, and not all Wasm binaries provide the corresponding source code.
Compared to Manticore, {\tool} not only provides interfaces and Wasm-tailored methods, but also allows symbol initialization through simple command line options.
Both the design and usage of {\tool} improve the usability.

\subsection{Quantitative Analysis}
Table~\ref{table:btree} illustrates the ratio of execution time between Manticore and {\tool} on $B_1$.
As we can see, {\tool} consistently outperforms Manticore, with at least 1.17x faster. In the most optimistic scenario, {\tool} can accelerate the analysis by nearly 5x.
Though the first row indicates that the gap between them is diminishing, we believe that manipulating a b-tree with degree as one only represents the most trivial situation. The remaining rows show that {\tool} can perform consistently on cases with more complex logics.

\begin{table}[t]
\caption{Ratio of time by Manticore and {\tool} on $B_1$.}
\vspace{-0.15in}
\centering
\resizebox{0.85\columnwidth}{!}{
\begin{tabular}{@{}lccccccccc@{}}
\toprule
                                                  % & \textbf{}  & \multicolumn{8}{c}{\textbf{\#elements}}                                                               \\
                                                \textbf{\#elements}  &   & \textbf{2} & \textbf{3} & \textbf{4} & \textbf{5} & \textbf{6} & \textbf{7} & \textbf{8} & \textbf{9} \\ \midrule
\multirow{3}{*}{\textbf{degree}} & \textbf{1} & 4.78       & 3.06       & 2.22       & 2.07       & 1.67       & 1.29       & 1.28       & 1.17       \\
                                                  & \textbf{2} & 3.10       & 3.32       & 3.14       & 3.63       & 2.68       & 2.86       & 2.64       & 2.64       \\
                                                  & \textbf{3} & 4.22       & 3.63       & 3.13       & 3.02       & 3.16       & 2.84       & 3.09       & 2.82       \\ \bottomrule
\end{tabular}
}
\label{table:btree}
\vspace{-0.2in}
\end{table}

\begin{table}[t]
\caption{Time used (s) by Manticore ($T_M$) and {\tool} ($T_S$) on $B_2$, and the number of cases (\#case) and instructions (\#ins).}
\vspace{-0.15in}
\centering
\resizebox{0.85\columnwidth}{!}{
\begin{tabular}{r|ccrrr}
\toprule
                      & \textbf{\#case} & \textbf{\#ins/\#Path}  & \multicolumn{1}{c}{\textbf{$T_M$}}      & \multicolumn{1}{c}{\textbf{$T_S$}}    & \multicolumn{1}{c}{\textbf{$T_M$/$T_S$}} \\
\midrule
\textbf{queue}        & 4               & 88K/19                & 13.33             & 2.86            & \textbf{4.67}              \\
\textbf{slist}        & 37              & 788K/216              & 256.82            & 56.10           & \textbf{4.58}              \\
\textbf{deque}        & 34              & 762K/125              & 82.81             & 17.30           & \textbf{4.79}              \\
\textbf{treeset}      & 6               & 138K/6                & 45.93             & 6.13            & \textbf{7.49}              \\
\textbf{stack}        & 2               & 44K/2                 & 1.04              & 0.72            & \textbf{1.44}              \\
\textbf{treetable}    & 13              & 297K/86               & 69.21             & 11.24           & \textbf{6.16}              \\
\textbf{ring}         & 3               & 63K/3                 & 2.59              & 1.24            & \textbf{2.09}              \\
\textbf{pqueue}       & 2               & 44K/28                & 20.34             & 2.46            & \textbf{8.27}              \\
\textbf{list}         & 37              & 807K/275              & 299.29            & 57.67           & \textbf{5.19}              \\
\textbf{array}        & 21              & 454K/260              & 221.66            & 20.68           & \textbf{10.72}             \\
\midrule
\textbf{Total}        & \textbf{159}    & \textbf{3,485K/1,020} & \textbf{1,013.01} & \textbf{176.39} & \textbf{5.74}    \\
\bottomrule
\end{tabular}
}
\label{table:gillian}
\vspace{-0.2in}
\end{table}

Table~\ref{table:gillian} illustrates their performance on $B_2$.
As we can see, {\tool} still outperforms Manticore. Under the test suite for the array structure, {\tool} can analyze a case in less than a second on average, while Manticore takes more than 10 seconds for each of them.
On average, {\tool} achieves nearly a six-fold improvement in efficiency.
We have to clarify that 159 cases analyzed by Manticore are simpler than the ones of {\tool}. This is because we have to manually delete \texttt{scanf}, \texttt{printf}, and \texttt{assert} such complex functions that manipulate the memory area and invoke imported ones, and add simpler but semantic-equivalent ones, \textit{i.e.}, \texttt{getchar}, \texttt{putchar}, and \texttt{if-else} statements, respectively.

Note that, in both tables, we measure the \#Path, \textit{i.e.}, the number of feasible paths. As we do not set a hard timeout limitation, both Manticore and {\tool} finish all cases with identical number of states.
It not only shows calculating path coverage is meaningless on $B_1$ and $B_2$, but also proves the soundness of {\tool}.

\subsection{Vulnerability Detection}
\textsc{Eunomia}~\cite{he2023eunomia} and \textsc{SymGX}~\cite{wang2023symgx} have extended {\tool} to detect more than 30 0-day vulnerabilities or security issues in C, Go, and SGX projects after compiling them to Wasm binaries. It proves the scalability and vulnerability detecting ability of {\tool}.

\section{Conclusion}
In this paper, we present {\tool}, an efficient and fully-functional symbolic execution engine for WebAssembly binaries.
{\tool} is designed to support full Wasm language features, including Wasm instruction emulation and external dependency modeling. We also introduced several techniques to improve its efficiency, such as the Wasm-specific linear memory modeling method.
Compared to Manticore, {\tool} illustrates its advantages, \textit{i.e.}, finishing the evaluation without additional human intervention while achieving 2 to 6 times efficiency improvement.

\section*{Acknowledgements}
This work was partly supported by the National Key R\&D of China (2022YFB4501802) and the National Natural Science Foundation of China (62141208). Yao Guo is the corresponding author.

\section*{Tool Availability}
The code of {\tool} is available at: \href{https://github.com/PKU-ASAL/SeeWasm}{https://github.com/PKU-ASAL/\-SeeWasm} and \href{https://doi.org/10.5281/zenodo.12671239}{https://doi.org/10.5281/zenodo.1267\-1239}.
The demonstration video is available at: \href{https://youtu.be/TgCBI3liKts}{https://youtu.be/TgCBI3liKts}.

\bibliographystyle{ACM-Reference-Format}
\bibliography{citation.bib}


\begin{thebibliography}{19}

%%% the Web address.

\ifx \showCODEN    \undefined \def \showCODEN     #1{\unskip}     \fi
\ifx \showDOI      \undefined \def \showDOI       #1{#1}\fi
\ifx \showISBNx    \undefined \def \showISBNx     #1{\unskip}     \fi
\ifx \showISBNxiii \undefined \def \showISBNxiii  #1{\unskip}     \fi
\ifx \showISSN     \undefined \def \showISSN      #1{\unskip}     \fi
\ifx \showLCCN     \undefined \def \showLCCN      #1{\unskip}     \fi
\ifx \shownote     \undefined \def \shownote      #1{#1}          \fi
\ifx \showarticletitle \undefined \def \showarticletitle #1{#1}   \fi
\ifx \showURL      \undefined \def \showURL       {\relax}        \fi
% invisible to TeX
\providecommand\bibfield[2]{#2}
\providecommand\bibinfo[2]{#2}
\providecommand\natexlab[1]{#1}
\providecommand\showeprint[2][]{arXiv:#2}

\bibitem[{ Srđan Pani{\'c}}(2024)]%
        {collections-c}
\bibfield{author}{\bibinfo{person}{{ Srđan Pani{\'c}}}.} \bibinfo{year}{2024}\natexlab{}.
\newblock \bibinfo{title}{{Repo of Collections C}}.
\newblock
\newblock
\urldef\tempurl%
\url{https://github.com/srdja/Collections-C}
\showURL{%
\tempurl}


\bibitem[{ Trail of Bits}(2024)]%
        {manticore}
\bibfield{author}{\bibinfo{person}{{ Trail of Bits}}.} \bibinfo{year}{2024}\natexlab{}.
\newblock \bibinfo{title}{{Repo of Manticore}}.
\newblock
\newblock
\urldef\tempurl%
\url{https://github.com/trailofbits/manticore}
\showURL{%
\tempurl}


\bibitem[Akinyemi(2024)]%
        {40lang}
\bibfield{author}{\bibinfo{person}{Stephen Akinyemi}.} \bibinfo{year}{2024}\natexlab{}.
\newblock \bibinfo{title}{Languages that can compile to {Wasm}}.
\newblock
\newblock
\urldef\tempurl%
\url{https://github.com/appcypher/awesome-wasm-langs}
\showURL{%
\tempurl}


\bibitem[Cadar et~al\mbox{.}(2008)]%
        {cadar2008klee}
\bibfield{author}{\bibinfo{person}{Cristian Cadar}, \bibinfo{person}{Daniel Dunbar}, \bibinfo{person}{Dawson~R Engler}, {et~al\mbox{.}}} \bibinfo{year}{2008}\natexlab{}.
\newblock \showarticletitle{Klee: unassisted and automatic generation of high-coverage tests for complex systems programs.}. In \bibinfo{booktitle}{\emph{OSDI}}, Vol.~\bibinfo{volume}{8}. \bibinfo{pages}{209--224}.
\newblock


\bibitem[Fragoso~Santos et~al\mbox{.}(2020)]%
        {fragoso2020gillian}
\bibfield{author}{\bibinfo{person}{Jos{\'e} Fragoso~Santos}, \bibinfo{person}{Petar Maksimovi{\'c}}, \bibinfo{person}{Sacha-{\'E}lie Ayoun}, {and} \bibinfo{person}{Philippa Gardner}.} \bibinfo{year}{2020}\natexlab{}.
\newblock \showarticletitle{Gillian, part i: a multi-language platform for symbolic execution}. In \bibinfo{booktitle}{\emph{Proceedings of the 41st ACM SIGPLAN Conference on Programming Language Design and Implementation}}. \bibinfo{pages}{927--942}.
\newblock


\bibitem[Haas et~al\mbox{.}(2017)]%
        {haas2017bringing}
\bibfield{author}{\bibinfo{person}{Andreas Haas}, \bibinfo{person}{Andreas Rossberg}, \bibinfo{person}{Derek~L Schuff}, \bibinfo{person}{Ben~L Titzer}, \bibinfo{person}{Michael Holman}, \bibinfo{person}{Dan Gohman}, \bibinfo{person}{Luke Wagner}, \bibinfo{person}{Alon Zakai}, {and} \bibinfo{person}{JF Bastien}.} \bibinfo{year}{2017}\natexlab{}.
\newblock \showarticletitle{Bringing the web up to speed with WebAssembly}. In \bibinfo{booktitle}{\emph{Proceedings of the 38th ACM SIGPLAN Conference on Programming Language Design and Implementation}}. \bibinfo{pages}{185--200}.
\newblock


\bibitem[He et~al\mbox{.}(2021)]%
        {he2021eosafe}
\bibfield{author}{\bibinfo{person}{Ningyu He}, \bibinfo{person}{Ruiyi Zhang}, \bibinfo{person}{Haoyu Wang}, \bibinfo{person}{Lei Wu}, \bibinfo{person}{Xiapu Luo}, \bibinfo{person}{Yao Guo}, \bibinfo{person}{Ting Yu}, {and} \bibinfo{person}{Xuxian Jiang}.} \bibinfo{year}{2021}\natexlab{}.
\newblock \showarticletitle{$\{$EOSAFE$\}$: security analysis of $\{$EOSIO$\}$ smart contracts}. In \bibinfo{booktitle}{\emph{30th USENIX security symposium (USENIX Security 21)}}. \bibinfo{pages}{1271--1288}.
\newblock


\bibitem[He et~al\mbox{.}(2023)]%
        {he2023eunomia}
\bibfield{author}{\bibinfo{person}{Ningyu He}, \bibinfo{person}{Zhehao Zhao}, \bibinfo{person}{Jikai Wang}, \bibinfo{person}{Yubin Hu}, \bibinfo{person}{Shengjian Guo}, \bibinfo{person}{Haoyu Wang}, \bibinfo{person}{Guangtai Liang}, \bibinfo{person}{Ding Li}, \bibinfo{person}{Xiangqun Chen}, {and} \bibinfo{person}{Yao Guo}.} \bibinfo{year}{2023}\natexlab{}.
\newblock \showarticletitle{Eunomia: enabling user-specified fine-grained search in symbolically executing WebAssembly binaries}. In \bibinfo{booktitle}{\emph{Proceedings of the 32nd ACM SIGSOFT International Symposium on Software Testing and Analysis}}. \bibinfo{pages}{385--397}.
\newblock


\bibitem[Jiang et~al\mbox{.}(2021)]%
        {jiang2021wana}
\bibfield{author}{\bibinfo{person}{Bo Jiang}, \bibinfo{person}{Yifei Chen}, \bibinfo{person}{Dong Wang}, \bibinfo{person}{Imran Ashraf}, {and} \bibinfo{person}{WK Chan}.} \bibinfo{year}{2021}\natexlab{}.
\newblock \showarticletitle{WANA: Symbolic execution of wasm bytecode for extensible smart contract vulnerability detection}. In \bibinfo{booktitle}{\emph{2021 IEEE 21st International Conference on Software Quality, Reliability and Security (QRS)}}. IEEE, \bibinfo{pages}{926--937}.
\newblock


\bibitem[Lehmann et~al\mbox{.}(2020)]%
        {lehmann2020everything}
\bibfield{author}{\bibinfo{person}{Daniel Lehmann}, \bibinfo{person}{Johannes Kinder}, {and} \bibinfo{person}{Michael Pradel}.} \bibinfo{year}{2020}\natexlab{}.
\newblock \showarticletitle{Everything old is new again: Binary security of $\{$WebAssembly$\}$}. In \bibinfo{booktitle}{\emph{29th USENIX Security Symposium (USENIX Security 20)}}. \bibinfo{pages}{217--234}.
\newblock


\bibitem[Marques et~al\mbox{.}(2022)]%
        {marques2022concolic}
\bibfield{author}{\bibinfo{person}{Filipe Marques}, \bibinfo{person}{Jos{\'e} Fragoso~Santos}, \bibinfo{person}{Nuno Santos}, {and} \bibinfo{person}{Pedro Ad{\~a}o}.} \bibinfo{year}{2022}\natexlab{}.
\newblock \showarticletitle{Concolic execution for webassembly}. In \bibinfo{booktitle}{\emph{36th European Conference on Object-Oriented Programming (ECOOP 2022)}}. Schloss-Dagstuhl-Leibniz Zentrum f{\"u}r Informatik.
\newblock


\bibitem[Mossberg et~al\mbox{.}(2019)]%
        {mossberg2019manticore}
\bibfield{author}{\bibinfo{person}{Mark Mossberg}, \bibinfo{person}{Felipe Manzano}, \bibinfo{person}{Eric Hennenfent}, \bibinfo{person}{Alex Groce}, \bibinfo{person}{Gustavo Grieco}, \bibinfo{person}{Josselin Feist}, \bibinfo{person}{Trent Brunson}, {and} \bibinfo{person}{Artem Dinaburg}.} \bibinfo{year}{2019}\natexlab{}.
\newblock \showarticletitle{Manticore: A user-friendly symbolic execution framework for binaries and smart contracts}. In \bibinfo{booktitle}{\emph{2019 34th IEEE/ACM International Conference on Automated Software Engineering (ASE)}}. IEEE, \bibinfo{pages}{1186--1189}.
\newblock


\bibitem[{SQLite}(2024)]%
        {sqlite}
\bibfield{author}{\bibinfo{person}{{SQLite}}.} \bibinfo{year}{2024}\natexlab{}.
\newblock \bibinfo{title}{{Wasm compatible SQLite}}.
\newblock
\newblock
\urldef\tempurl%
\url{https://sqlite.org/wasm/doc/trunk/index.md}
\showURL{%
\tempurl}


\bibitem[{The Gillian Platform}(2024)]%
        {gillian}
\bibfield{author}{\bibinfo{person}{{The Gillian Platform}}.} \bibinfo{year}{2024}\natexlab{}.
\newblock \bibinfo{title}{{Repo of Gillian Collections C}}.
\newblock
\newblock
\urldef\tempurl%
\url{https://github.com/GillianPlatform/collections-c-for-gillian}
\showURL{%
\tempurl}


\bibitem[Wang et~al\mbox{.}(2023)]%
        {wang2023symgx}
\bibfield{author}{\bibinfo{person}{Yuanpeng Wang}, \bibinfo{person}{Ziqi Zhang}, \bibinfo{person}{Ningyu He}, \bibinfo{person}{Zhineng Zhong}, \bibinfo{person}{Shengjian Guo}, \bibinfo{person}{Qinkun Bao}, \bibinfo{person}{Ding Li}, \bibinfo{person}{Yao Guo}, {and} \bibinfo{person}{Xiangqun Chen}.} \bibinfo{year}{2023}\natexlab{}.
\newblock \showarticletitle{Symgx: Detecting cross-boundary pointer vulnerabilities of sgx applications via static symbolic execution}. In \bibinfo{booktitle}{\emph{Proceedings of the 2023 ACM SIGSAC Conference on Computer and Communications Security}}. \bibinfo{pages}{2710--2724}.
\newblock


\bibitem[{WASI}(2022)]%
        {wasi}
\bibfield{author}{\bibinfo{person}{{WASI}}.} \bibinfo{year}{2022}\natexlab{}.
\newblock \bibinfo{title}{{WebAssembly System Interface}}.
\newblock
\newblock
\urldef\tempurl%
\url{https://wasi.dev/}
\showURL{%
\tempurl}


\bibitem[{WebAssembly}(2021)]%
        {wasi-sdk}
\bibfield{author}{\bibinfo{person}{{WebAssembly}}.} \bibinfo{year}{2021}\natexlab{}.
\newblock \bibinfo{title}{{WASI SDK 14}}.
\newblock
\newblock
\urldef\tempurl%
\url{https://github.com/WebAssembly/wasi-sdk/releases/tag/wasi-sdk-14}
\showURL{%
\tempurl}


\bibitem[{WebAssembly Community Group}(2022)]%
        {wasm-specification}
\bibfield{author}{\bibinfo{person}{{WebAssembly Community Group}}.} \bibinfo{year}{2022}\natexlab{}.
\newblock \bibinfo{title}{{WebAssembly} spcification}.
\newblock
\newblock
\urldef\tempurl%
\url{https://webassembly.github.io/spec/core/}
\showURL{%
\tempurl}


\bibitem[ymc9(2023)]%
        {wasm-projects}
\bibfield{author}{\bibinfo{person}{ymc9}.} \bibinfo{year}{2023}\natexlab{}.
\newblock \bibinfo{title}{Projects that use {Wasm}}.
\newblock
\newblock
\urldef\tempurl%
\url{https://dev.to/zenstack/what-are-people-building-with-webassembly-2eh4}
\showURL{%
\tempurl}


\end{thebibliography}

% \appendix

\end{document}